\documentclass[final,3p,times,twocolumn]{elsarticle}
\usepackage{amssymb}

\usepackage{hyperref}
\usepackage{color}
\newcommand{\ehm}{\scalebox{0.6}{\rm EHM}}
\newcommand{\cdw}{\scalebox{0.6}{\rm CDW}}
\newcommand{\rb}[1]{\raisebox{1.5ex}[-1.5ex]{#1}}

\journal{Physica B}
\begin{document}
\begin{frontmatter}

\title{
Critical behavior  of the extended Hubbard model with bond dimerization
}

\author[HGW]{Satoshi Ejima}
\author[HGW,RIKEN]{Florian Lange}
\author[Oxford]{Fabian H. L. Essler}
\author[HGW]{Holger Fehske}

\address[HGW]{
Institut f\"ur Physik, 
Ernst-Moritz-Arndt-Universit\"at Greifswald, 
17487 Greifswald, Germany
}
\address[RIKEN]{
Computational Condensed Matter Physics Laboratory, 
RIKEN, Wako, Saitama 351-0198, Japan
}
\address[Oxford]{
 The Rudolf Peierls Centre for Theoretical Physics, 
 Oxford University, 
 Oxford OX1 3NP, 
 United Kingdom
}

\begin{abstract}
Exploiting the matrix-product-state based density-matrix 
renormalization group (DMRG) technique we study  the one-dimensional 
extended ($U$-$V$) Hubbard model with explicit bond dimerization
in the half-filled band sector.  In particular we investigate the nature of the quantum phase transition, taking place with  growing ratio $V/U$ 
between the symmetry-protected-topological and charge-density-wave insulating states.  
The (weak-coupling) critical line of continuous Ising transitions with central charge $c=1/2$ terminates at a tricritical point 
belonging to the universality class of the dilute Ising model 
with $c=7/10$. We demonstrate that our DMRG data perfectly match with (tricritical) Ising exponents, e.g., for the order parameter $\beta=1/8$ 
(1/24) and correlation length $\nu=1$ (5/9). Beyond the  tricritical Ising point, in the strong-coupling regime,  the quantum phase transition becomes first order. 
\end{abstract}


\end{frontmatter}
\section{Introduction}
Half a century has passed since it was proposed, yet the Hubbard model~\cite{Hubbard1963} is still a key Hamiltonian for the investigation 
of strongly correlated electron systems.   Originally designed to describe the ferromagnetism of transition metals, in successive studies the 
Hubbard model has also been used for heavy fermions and high-temperature superconductors.  The physics of the model is governed by the competition 
between the itinerancy of the charge carriers and their local Coulomb interaction.  In one dimension (1D), seen from a theoretical point of view,
the Hubbard model is a good starting point to explore, for example,
Tomonaga-Luttinger liquid behavior (including spin-charge separation). 

While the 1D Hubbard model is exactly solvable by Bethe Ansatz~\cite{EFGKK05}, most of its extensions are no longer
integrable. This is even true if only the Coulomb interaction between electrons on nearest-neighbor lattice sites is added. 
The ground-state phase diagram of this so-called extended Hubbard model (EHM) is still a hotly debated issue.  
At half filling, this relates in particular to the recently discovered bond-order-wave (BOW) state located in between 
spin-density-wave (SDW) and charge-density-wave (CDW)
phases~\cite{Na99,Na2000}. To characterize the BOW state and determine its 
phase boundaries considerable efforts were undertaken in the last few years, using both analytical~\cite{TF02,TF04} and numerical~\cite{Je02,SBC04,EN07} 
methods.

At present, quantum phase transitions between topologically trivial and
nontrivial  states arouse great interest~\cite{GW09,PTBO10,PBTO12}. 
In this context, extensions of the half-filled EHM also attracted
attention, mainly  with regard to the formation of
symmetry-protected-topological (SPT) states~\cite{PTBO10}.
Including an alternating ferromagnetic spin interaction~\cite{LEF15} or an explicit dimerization~\cite{EELF16} in the EHM, 
the SDW and BOW phases are completely replaced by an SPT insulator,  whereby a quantum phase transition occurs between the SPT and 
the CDW, the area of which shrinks. Most interestingly, the SPT-CDW continuous Ising  transition with central charge $c=1/2$ ends at a tricritical point, 
belonging to the universality class of the tricritical Ising model,
a second minimal model with $c=7/10$~\cite{FQS84,FQS85}.  
Above this point, the quantum phase transition becomes first order. 
In Ref.~\cite{EELF16} it has been demonstrated that 
the transition region of the EHM with bond dimerization can be 
described by the triple sine-Gordon model by extending 
the former bosonization analysis~\cite{BEG06}. 
The predictions of field theory regarding  
power-law (exponential) decay of the density-density (spin-spin) and bond-order 
correlation functions are shown to be in excellent accordance with the numerical data 
obtained by a matrix-product-states (MPS) based density-matrix 
renormalization group (DMRG) technique~\cite{Wh92,Sch11}.

The Ising criticality of the EHM with explicit dimerization was established in early work~\cite{BEG06} that also specifies the critical exponents.
The critical exponents at the tricritical point should differ from those
at the ordinary Ising  transition because the tricritical Ising  quantum
phase transition belongs to a different universality class.

Simulating the neutral gap and the CDW order parameter by DMRG, 
in this paper we will determine the critical exponents at both Ising and 
tricritical Ising transitions. The paper is structured as follows. 
Section~\ref{sec:model} introduces the model Hamiltonians under consideration 
and discusses their ground-state properties. The critical exponents will be derived  in Sect.~\ref{sec:crit-expos}. Section~\ref{sec:summary}
summarizes our main results.

\section{Model}
\label{sec:model}

\subsection{Extended Hubbard model}

\begin{figure}[tb]
 \includegraphics[width=\columnwidth,clip]{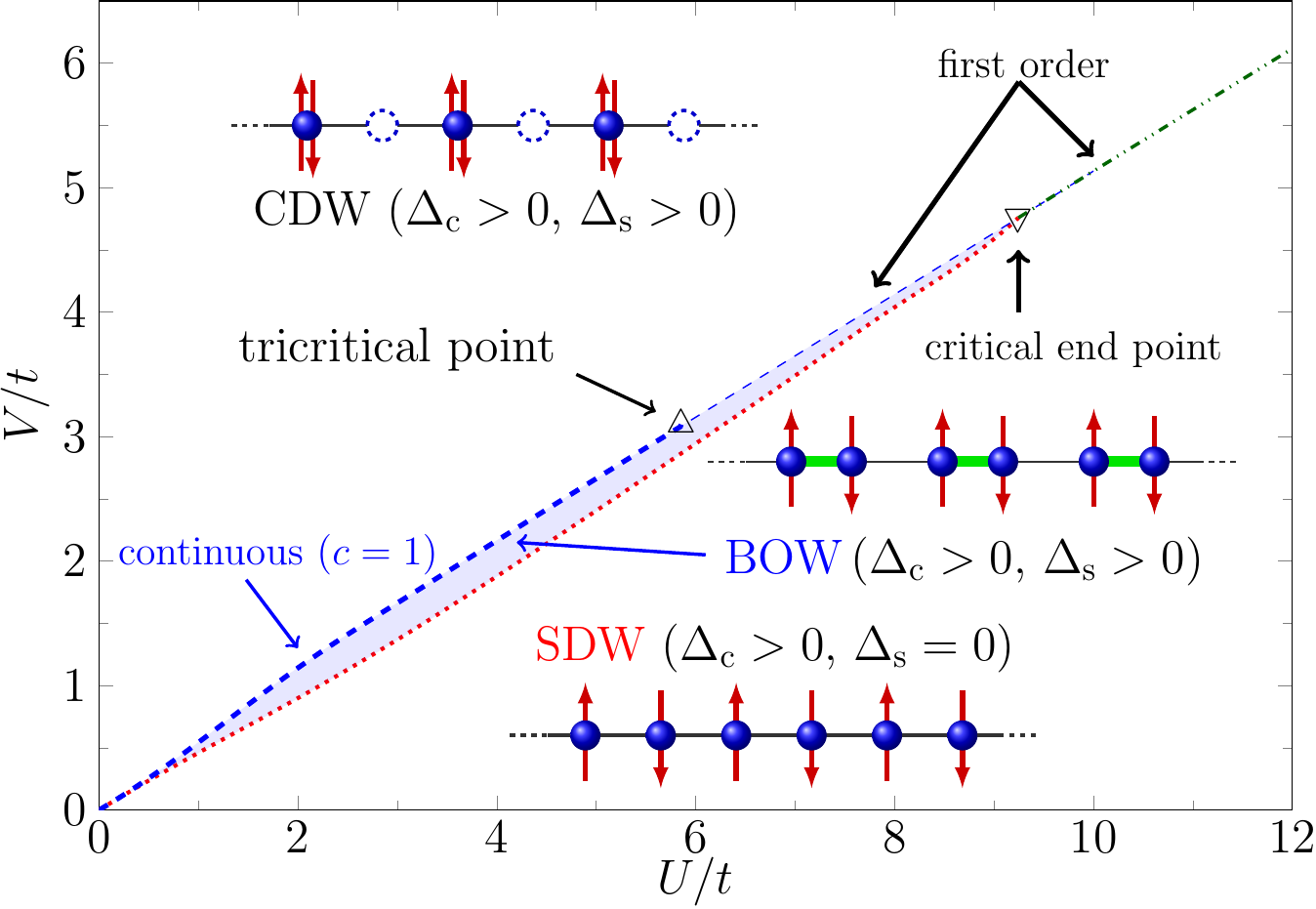}
 \caption{DMRG ground-state phase diagram of the 1D EHM~(\ref{H_EHM})
 at half filling~\cite{EN07}. The red dotted line gives the continuous
 SDW-BOW transition. The bold (thin) blue dashed line marks the
 continuous (first-order) BOW-CDW transition and the green dashed-dotted 
 line denotes the first-order SDW-CDW transition.
 }
 \label{PD-EHM}
\end{figure}

The Hamiltonian of the EHM is defined as
\begin{eqnarray}
\hat{H}_{\ehm} &=&
 -t \sum_{j\sigma}
 (\hat{c}^\dagger_{j\sigma}\hat{c}_{j+1\sigma}^{\phantom{\dagger}} 
 + {\rm H.c.}) 
 \nonumber\\
 &&+U \sum_{j}\left(\hat{n}_{j\uparrow}-\frac{1}{2}\right)
                \left(\hat{n}_{j\downarrow}-\frac{1}{2}\right)
 \nonumber\\
 &&+ V \sum_{j} (\hat{n}_{j}-1)
  (\hat{n}_{j+1}-1) \,,
  \label{H_EHM}
\end{eqnarray}
where $\hat{c}^\dagger_{j\sigma}$ ($\hat{c}^{\phantom{}}_{j\sigma}$)
creates (annihilates) an electron with spin projection 
$\sigma=\uparrow,\downarrow$ at Wannier site $j$, 
$\hat{n}_{j\sigma}=\hat{c}^\dagger_{j\sigma}\hat{c}^{\phantom{}}_{j\sigma}$,
and $\hat{n}_{j}=\hat{n}_{j\uparrow}+\hat{n}_{j\downarrow}$.
In the Hubbard model limit ($V=0$), at half-filling, no long-range order exists. 
Instead the system shows fluctuating SDW order. The spin (charge) 
excitations are gapless (gapped) $\forall U>0$~\cite{EFGKK05}. 
At finite $V$, for $V/U\lesssim1/2$, the ground state is still a SDW.
When  $V/U$ becomes larger than 1/2 a 2$k_{\scalebox{0.6}{\rm F}}$-CDW is formed. 
As pointed out first by Nakamura~\cite{Na99,Na2000} and confirmed
later by various analytical and numerical studies~\cite{SBC04,EN07,SSC02,TTC06}, 
the SDW and CDW phases are separated by a narrow BOW phase below the critical end point, 
$(U_{\rm ce}^{\ehm}$,$V_{\rm ce}^{\ehm})\approx (9.25t,4.76t)$. 
In the BOW phase  translational symmetry is spontaneously broken, 
which implies that the spin gap opens passing the SDW-BOW phase boundary
at fixed $U<U_{\rm ce}^{\ehm}$. Increasing $V$ further, the system enters the CDW phase 
with finite spin and charge gaps. The BOW-CDW Gaussian 
transition line with central charge $c=1$ terminates at the tricritical point, 
$(U_{\rm tr}^{\ehm}, V_{\rm tr}^{\ehm})\approx (5.89t, 3.10t)$~\cite{EN07}. 
For $U_{\rm tr}^{\ehm}<U<U_{\rm ce}^{\ehm}$,  the BOW-CDW transition becomes first order, 
characterized by  a jump in the spin gap (see, Fig.~3 in Ref.~\cite{EN07}).
Figure~\ref{PD-EHM} summarizes the rich physics of the half-filled EHM.

\begin{figure}[tb]
 \includegraphics[width=\columnwidth,clip]{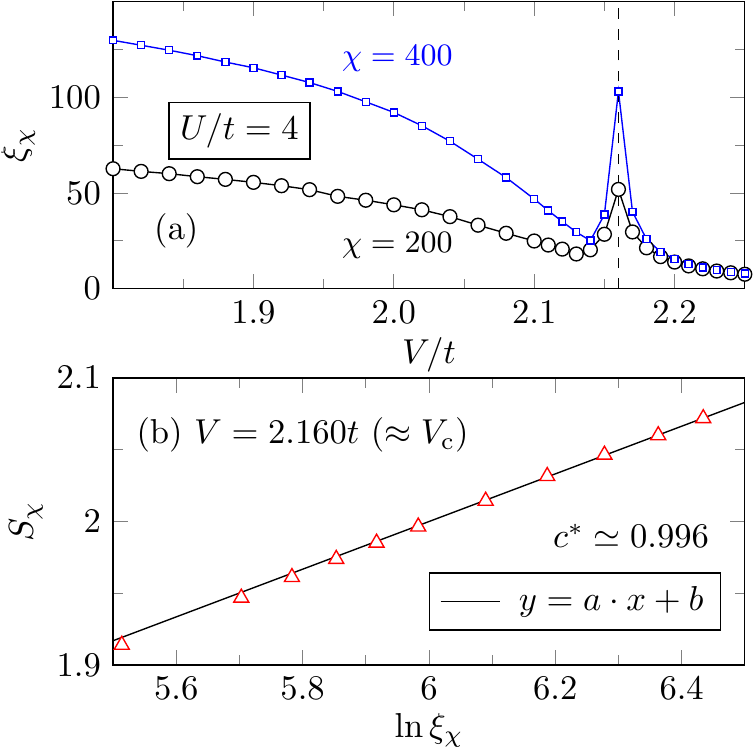}
 \caption{(a): Correlation length $\xi_\chi$ of the EHM 
 as a function of $V/t$ for $U/t=4$ obtained from iDMRG. 
 The dashed line indicates the BOW-CDW transition point.
 (b): von Neumann entropy $S_\chi$ as a function of logarithm 
 of $\xi_\chi$ at $V\approx V_{\rm c}$ for $U/t=4$. The iDMRG
 data for $\ln\xi_\chi>6$ ($\chi\geq 1800$) provide us the 
 numerically obtained central charge $c^\ast\simeq 0.996$
 by fitting to Eq.~(\ref{eq-vN}).
 }
 \label{xi-ehm}
\end{figure}

The criticality at the continuous BOW-CDW transition line can be 
verified numerically by extracting, e.g., the central charge 
from the the correlation length ($\xi_\chi$) and von Neumann entropy ($S_\chi$), 
where $\xi_\chi$ can be obtained from the second largest eigenvalue
of the transfer matrix for some bond dimension $\chi$ used in a 
infinite DMRG (iDMRG) simulation~\cite{Sch11,Mc08}. Conformal field
theory tells us that the von Neumann entropy for a system 
between two semi-infinite chains is~\cite{CC04} 
\begin{eqnarray}
 S_\chi=\frac{c}{6}\ln\xi_\chi +s_0 
 \label{eq-vN}
\end{eqnarray}
with a non-universal constant $s_0$. 

Figure~\ref{xi-ehm}(a) shows iDMRG results of $\xi_\chi$ 
as a function of $V/t$ for fixed $U/t=4$. Since the system 
is critical in the SDW phase and at the BOW-CDW transition point, 
we find a rapid increase of $\xi_\chi$ in the SDW
phase and a distinct peak at the BOW-CDW critical point 
($V_{\rm c}/t \approx 2.160$) when we increase $\chi$ from 200 to 400. 
This indicates the divergence of the correlation length 
$\xi_\chi\to\infty$ as $\chi\to\infty$. Now, plotting the 
von Neumann entropy $S_\chi$ as a function of $\ln\xi_\chi$ 
and fitting the graph to Eq.~(\ref{eq-vN}), the criticality 
at $V=V_{\rm c}$ can be proved, as demonstrated by
Fig.~\ref{xi-ehm}(b). The obtained $c^\ast\simeq 0.996$ for iDMRG data with 
$\chi\geq1800$ corroborates the Gaussian transition  resulting from a bosonization 
analysis~\cite{TF02,TF04}. Note that for the confirmation of the SDW-BOW
transition much larger bond dimensions $\chi$ are required in order to make clear  
the convergence of $\xi_\chi$ in the BOW phase of Fig.~\ref{xi-ehm}.

\subsection{EHM with explicit bond dimerization}
 
\begin{figure}[tb]
 \includegraphics[width=\columnwidth,clip]{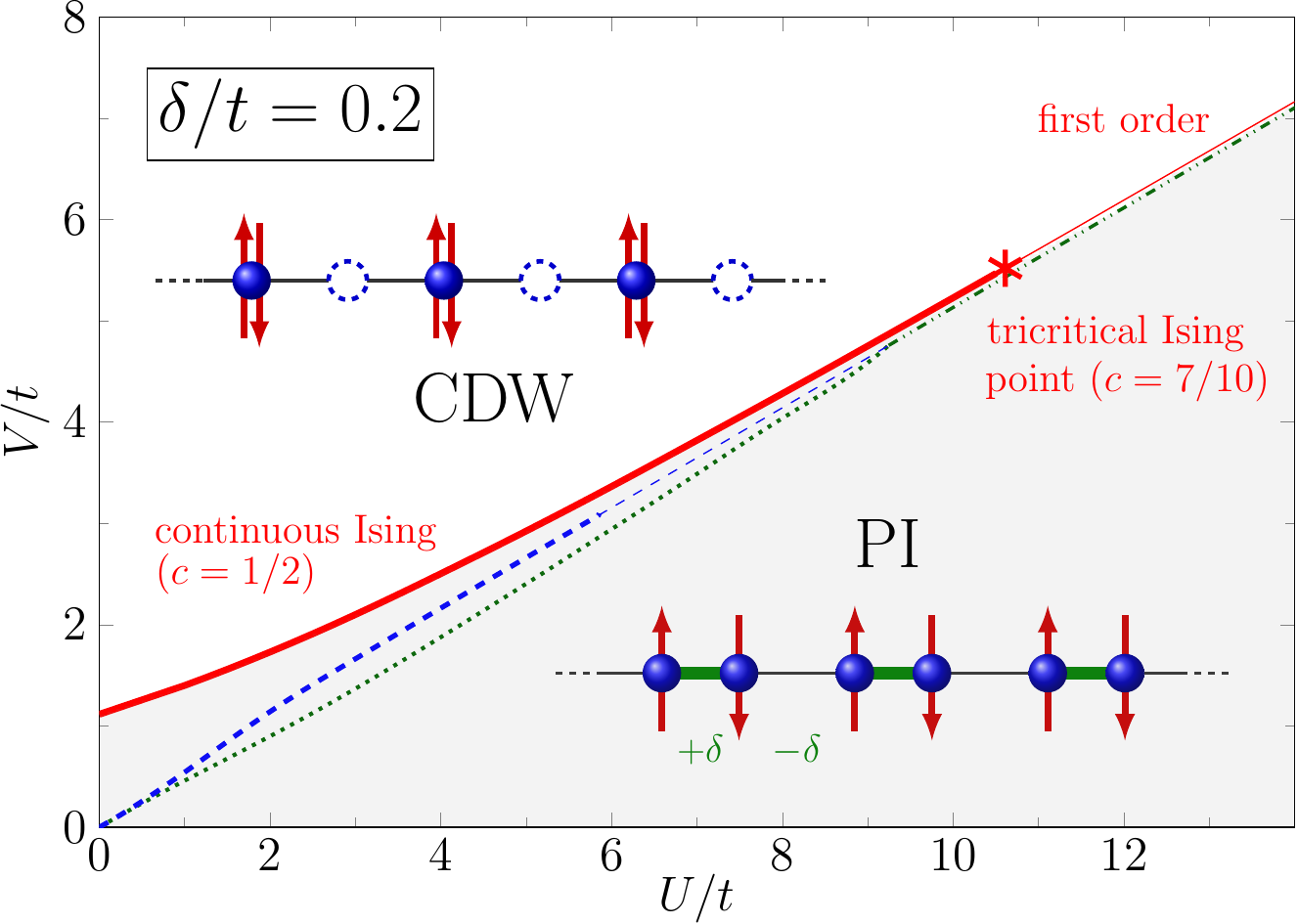}
 \caption{Ground-state phase diagram of the 1D EHM
 with bond dimerization in the half-filled band 
 sector~\cite{EELF16}.
 The red solid line marks the PI-CDW phase boundaries 
 for $\delta/t=0.2$. The tricritical Ising point 
 [$U_{\rm tr}$, $V_{\rm tr}$] separates continuous Ising
 and first-order phase transitions. For comparisons, 
 the phase boundaries of the pure EHM ($\delta=0$) were included.
 }
 \label{PD-EPHM}
\end{figure}

Let us now add a staggered bond dimerization to 
the EHM, $\hat{H} = \hat{H}_{\ehm}+\hat{H}_{\delta}$, where
\begin{equation}
 \hat{H}_{\delta}= -t\sum_{j\sigma}\delta(-1)^j
 (\hat{c}^\dagger_{j\sigma}\hat{c}_{j+1\sigma}^{\phantom{\dagger}}
 + {\rm H.c.})\, .
 \label{H_EPHM}
\end{equation}
Previous studies of this model have shown that  
the low lying excitations in the large-$U$ limit are chargeless 
spin-triplet and spin-singlet excitations~\cite{GNT99,Gi03,GBSTK97,NF81,Ts92,US96,ETD97}, 
whereby the dynamics is described by an effective spin-Peierls Hamiltonian.
Moreover, at  finite $U$,  the Tomonaga-Luttinger parameters have been explored at and near commensurate fillings 
by DMRG~\cite{EGN06}. Particularly for half filling, it has been proven by perturbative~\cite{GGR05,DM05} and renormalization 
group~\cite{TF04,SS02,TO02} approaches that the system realizes  
Peierls insulator (PI) and CDW phases in the weak-coupling regime.   
According to weak-coupling renormalization-group results~\cite{TF04}, 
any finite bond dimerization $\delta$ will change the universality class of the continuous BOW-CDW transition (realized in the pure EHM) 
from Gaussian to Ising type. Thereby the PI-CDW transition in the weak-to-intermediate
coupling regime belongs to the universality class of the two-dimensional (2D)
Ising model~\cite{TF04,BEG06}. 

Even more interesting physics appears analyzing the
intermediate-to-strong-coupling regime~\cite{EELF16} by analogy with  
an effective spin-1 (EHM) system with alternating ferromagnetic
spin interaction~\cite{LEF15}: Here the continuous PI-CDW Ising transition line
with central charge $c=1/2$ terminates at a tricritical point that
belongs to the universality class of the 2D dilute Ising model with $c=7/10$. 
Above the tricritical Ising point the quantum phase transition becomes first order.
Displaying the ground-state phase diagram,  
Fig.~\ref{PD-EPHM} summarizes these results.
A field theoretical description of the tricritical transition region has been performed 
in terms of a triple sine-Gordon model~\cite{EELF16}, based on the bosonization analysis in Ref.~\cite{BEG06}, 
providing  results for the decay of various correlation functions, such as the density-density, bond-order or spin-spin two-points functions.
The predictions of field theory are in excellent agreement with iDMRG data.

\section{Critical exponents}
\label{sec:crit-expos}
In the following, we give further evidence for the Ising respectively the 
tricritical Ising universality classes of the quantum phase transitions
in the EHM with bond dimerization by calculating the critical exponents 
of various physical quantities.  When approaching a continuous phase 
transition by varying a parameter  (e.g., a coupling strength) 
$g$ of the Hamiltonian, the correlation length diverges as
\begin{eqnarray}
 \xi \propto \left|g-g_{\rm c}\right|^{-\nu}\, .
\end{eqnarray}
Here, $g_{\rm c}$ denotes the (critical) value of $g$ at the transition 
point and $\nu$ is the corresponding critical exponent.  
Other quantities such as the order parameters or energy gaps also show 
power-law behavior. In this way the system is characterized 
by a set of universal exponents near the continuous phase transitions. 
The exact values of the most common exponents for the 2D Ising and 
tricritical Ising universality classes are listed in Table~\ref{crit_expo}.
\begin{table}[t]
 \begin{center}
 \begin{tabular}{c|c|c|c}
  \hline
 &  &  & tricritical  \\ 
 \rb{quantity}      &  \rb{exponent}    &  \rb{Ising} &  Ising \\ \hline \hline
  magnetization     & $\beta$           &  1/8        &        1/24  \\
 correlation length & $\nu$             &   1         &        5/9   \\
 pair correlation   & $\eta$            &  1/4        &        3/20  \\
  \hline
 \end{tabular} 
 \end{center}
 \caption{Critical exponents belonging to the Ising and tricritical 
 Ising universality classes in
 2D~\cite{LMC91,DiFrancesco1997,Mussardo2009}.
 The critical exponent $\eta$ for the pair correlation function 
 has been confirmed in Ref.~\cite{EELF16}.
 }
 \label{crit_expo}
\end{table}
The exponents satisfy the following scaling relation 
\begin{eqnarray}
 \frac{\nu}{2}(\eta+{\rm d}-2)=\beta\, , 
\end{eqnarray}
where d is the spatial dimension (in our case ${\rm d}=2$).

For the EHM with bond dimerization, $\beta$ and $\nu$
can be extracted from the CDW order parameter and 
the neutral gap, respectively. The CDW order parameter 
is defined as
\begin{eqnarray}
 m_{\cdw}=\frac{1}{L}\sum_j (-1)^j (\hat{n}_j-1)\,.
\end{eqnarray}
The neutral gap is obtained from 
\begin{eqnarray}
 \Delta_{\rm n}(L)=E_1(N)-E_0(N)\, ,
\end{eqnarray}
where $E_{0}(N)$ [$E_{1}(N)$] denotes the energy of the ground state
[first excited state]  of a system with $L$ sites, $N$
electrons, and vanishing total spin $z$ component.

\subsection{Ising transition}
\begin{figure}[tb]
 \includegraphics[width=\columnwidth,clip]{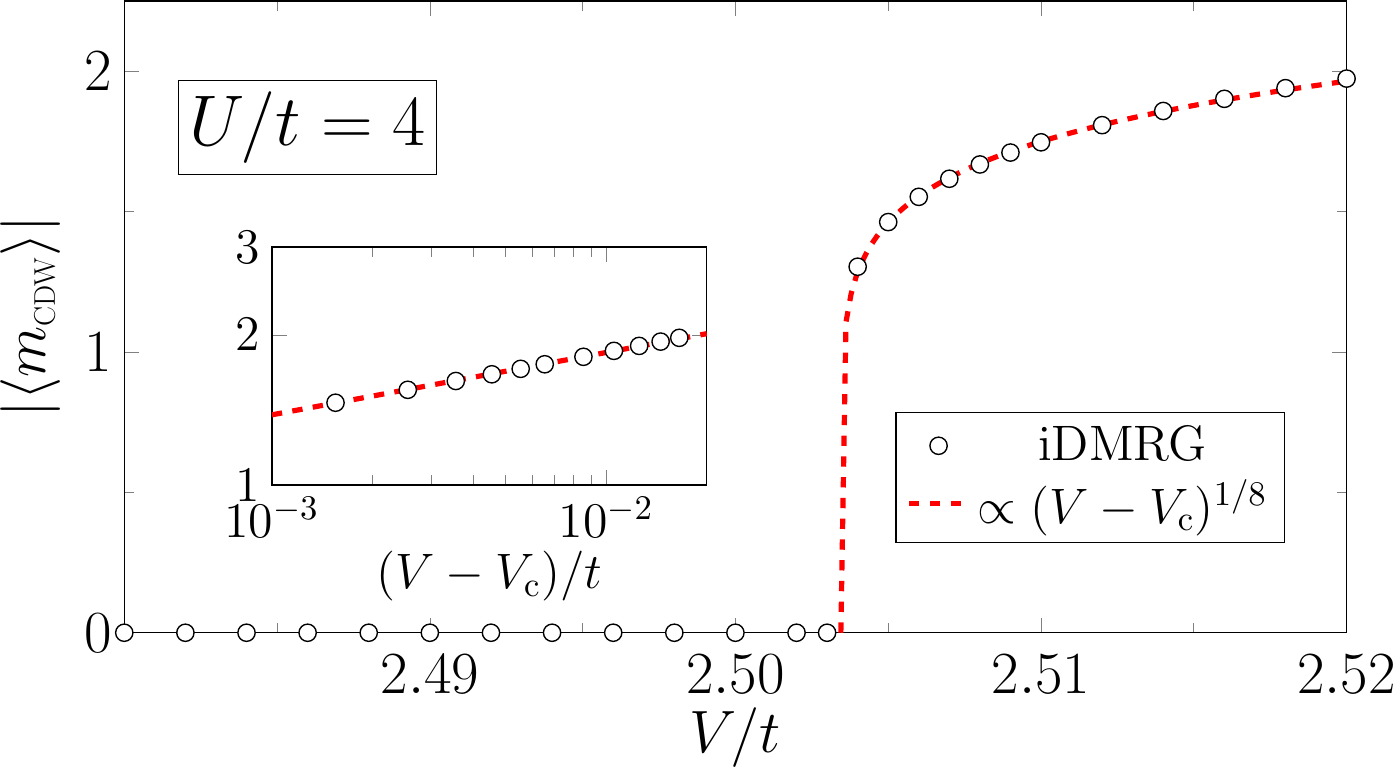}
 \caption{Absolute value of the CDW order parameter 
 in the vicinity of  the Ising transition at fixed $U/t=4$.
 Symbols are iDMRG data; the dashed line displays the fitting function 
 $|\langle m_{\rm CDW}\rangle|\propto(V-V_{\rm tr})^{\beta}$ 
 with critical exponent $\beta=1/8$ (Ising universality class).
 Inset: Log-log plot of the order parameter for $V>V_{\rm tr}$ 
 demonstrating the power-law decay with exponent $\beta$.}
 \label{Ising-beta}
\end{figure}
We now show that the critical exponents $\beta=1/8$ and $\nu=1$ 
follow from (i)DMRG simulations by varying $V$ at fixed $U$ and $\delta$,
just as the corresponding phase transition line was obtained in Fig.~\ref{PD-EPHM}.
Note that $\beta=1/8$ and $\nu=1$ were extracted in Ref.~\cite{BEG06} by means 
of the DMRG method, varying $\delta$ for fixed $U$ and $V$.

Figure~\ref{Ising-beta} gives the CDW order parameter 
as a function of $V/t$, fixing $U/t=4$ and $\delta/t=0.2$, 
calculated by iDMRG technique with bond dimensions $\chi=800$. 
Obviously, in the CDW (PI) realized for $V>V_{\rm c}$ 
$(V<V_{\rm c})$,  $|m_{\cdw}|$ is finite (zero). 
Using $V_{\rm c}/t \approx 2.5035$, the iDMRG data are well fitted 
by  $(V-V_{\rm c})^{\,\beta}$ near the transition, where the critical exponent  $\beta=1/8$ can be easily read off from a log-log plot; see inset of Fig.~\ref{Ising-beta}.

\begin{figure}[tb]
 \includegraphics[width=\columnwidth,clip]{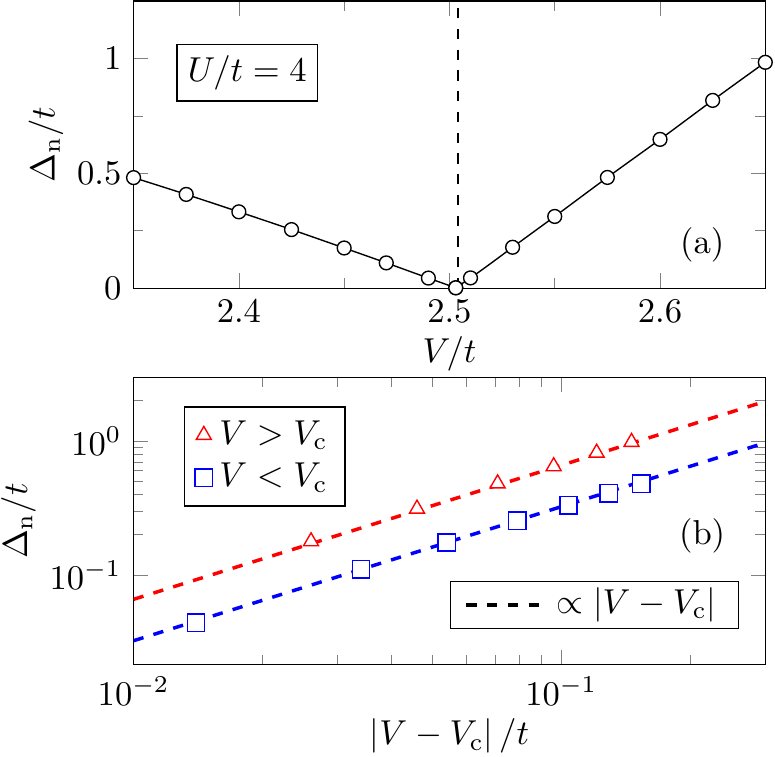}
 \caption{(a): Neutral gap $\Delta_{\rm n}$ near the 
 Ising transition at fixed $U/t=4$ (symbols are DMRG
data taken from Ref.~\cite{EELF16}).
 (b): Log-log plots of $\Delta_{\rm n}$ as a function of 
 $|V-V_{\rm c}|$, fitted by $|V-V_{\rm c}|^{\nu}$ with $\nu=1$ 
 (Ising universality class).
 }
 \label{Ising-nu}
\end{figure}

Extrapolating the values of the neutral gap $\Delta_{\rm n}$
to the thermodynamic limit, the critical exponent $\nu=1$ is  verified, 
as demonstrated by Fig.~\ref{Ising-nu}. Increasing $V$ at fixed $U/t=4$, 
the neutral gap decreases linearly and closes at the Ising transition point. 
If $V$ grows further, $\Delta_{\rm n}$ opens again with linear slope.
This is clearly visible in the log-log plots representation, both 
for $V>V_{\rm c}$ and $V<V_{\rm c}$; see Fig.~\ref{Ising-nu}(b).

\subsection{Perturbed tricritical Ising model}
\begin{figure}[t]
 \includegraphics[width=\columnwidth,clip]{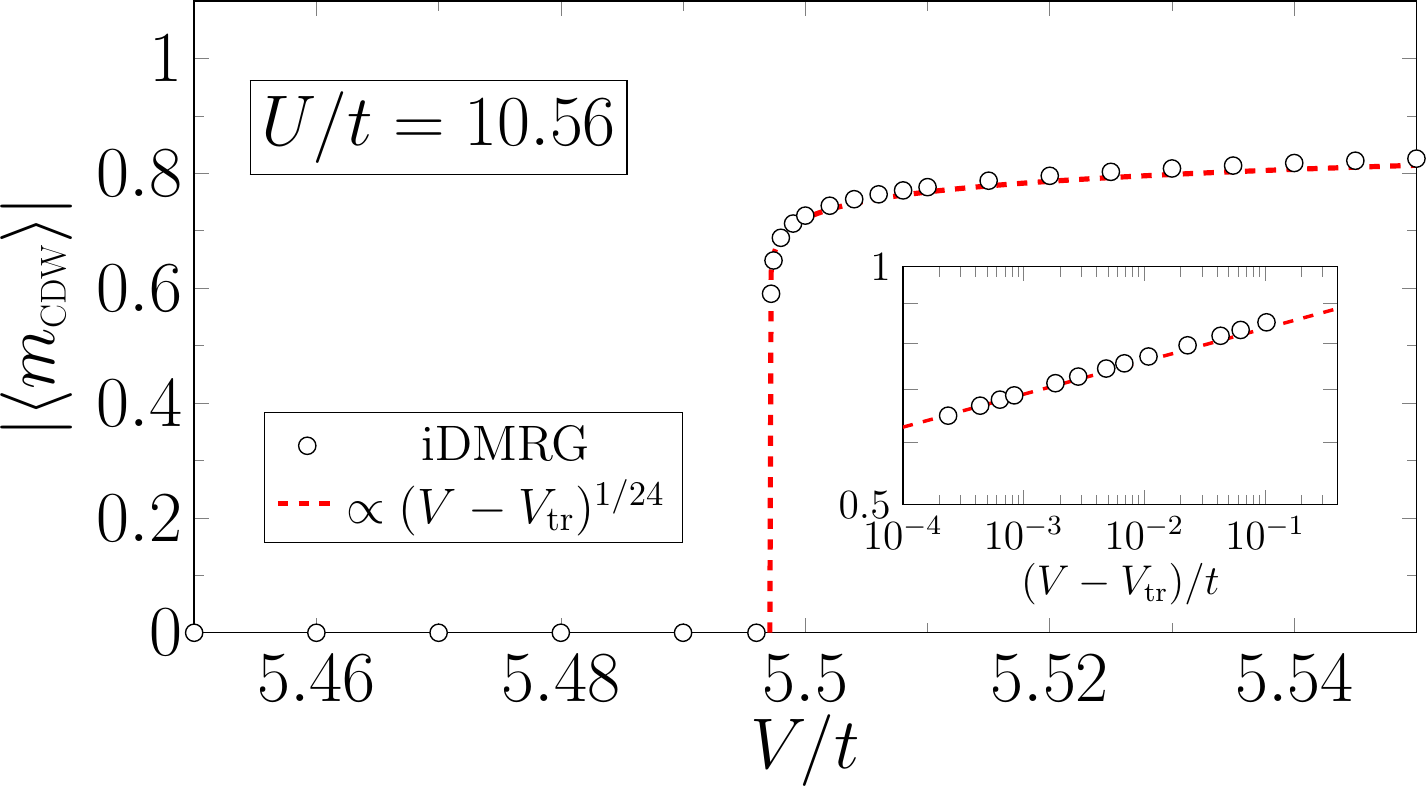}
 \caption{Absolute value of the CDW order parameter 
 in the vicinity of  the tricritical Ising point at fixed $U/t=10.56$.
 Symbols are iDMRG data; the dashed line displays the fitting function 
 $|\langle m_{\rm CDW}\rangle|\propto(V-V_{\rm tr})^{\beta}$ 
 with critical exponent $\beta=1/24$ (tricritical  Ising universality class).
 Inset: Log-log plot of the order parameter for $V>V_{\rm tr}$ 
 demonstrating the power-law decay with exponent $\beta$.
 }
 \label{TIM-beta}
\end{figure}

As quoted above and demonstrated in Ref.~\cite{EELF16}, the
tricritical point in the EHM with bond dimerization belongs to the
universality class of the 2D tricritical Ising model with the critical
exponents given in Table~\ref{crit_expo}.  Let us emphasize that it is 
exceptionally challenging to verify the critical exponents 
at the tricritical Ising point numerically, not least because 
one first has to determine the tricritical point itself, 
with high precision, varying $U$ and $V$ simultaneously~\cite{EELF16}. 

The exponent $\eta$ characterizes the power-law decay of the CDW
order-parameter two-point function at the critical point. As shown in
Ref.~\cite{EELF16} one has
\begin{equation}
\langle (-1)^\ell(\hat{n}_{j+\ell}-1)(\hat{n}_{j}-1)\rangle\propto
  \ell^{-3/20}\ ,\quad \ell\gg 1\ .
\end{equation}
This establishes that $\eta=3/20$. In order to determine the exponents
$\beta$ and $\nu$ one needs to consider the off-critical regime. We
therefore consider the perturbation of the tricritical Ising
conformal field theory by the ``energy operator'' $\epsilon(x)$, which has
conformal dimensions $\left(\Delta_\epsilon,\bar{\Delta}_\epsilon\right)=
\left(\frac{1}{10},\frac{1}{10}\right)$~\cite{LMC91,DiFrancesco1997,Mussardo2009}
\begin{equation}
H=H_{\rm CFT}+h\int dx\ \epsilon(x)\ .
\label{PCFT}
\end{equation}
The perturbing operator has scaling dimension $d=1/5$ and is therefore
relevant in the renormalization group (RG) sense. It generates a
spectral gap $M$ that scales as
\begin{equation}
M\sim Ch^{1/(2-d)}=Ch^{5/9},
\end{equation}
where $C$ is a constant. This identifies the critical exponent
$\nu=5/9$. The magnetization operator $\sigma(x)$ in the tricritical
Ising model has scaling dimension
$\left(\Delta_\sigma,\bar{\Delta}_\sigma\right)=\left(\frac{3}{80},\frac{3}{80}\right)$. 
In the perturbed theory (\ref{PCFT}) it acquires a non-zero 
expectation value that scales as
\begin{equation}
 \langle\sigma(x)\rangle\sim D
  h^{\Delta_\sigma/(1-\Delta_\epsilon)}
  =D h^{1/24}\ ,
\end{equation}
where $D$ is a constant. This identifies the critical exponent
$\beta=1/24$. 

\begin{figure}[tb]
 \includegraphics[width=\columnwidth,clip]{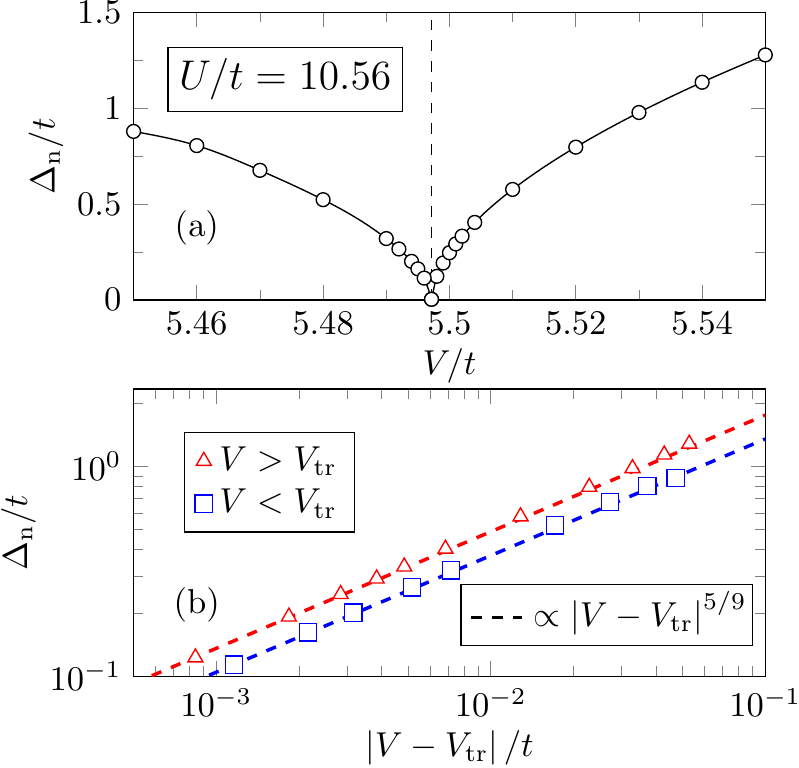}
 \caption{(a): DMRG data for the neutral gap $\Delta_{\rm n}$ 
 in the vicinity of the tricritical Ising point where $U/t=10.56$.
 (b): Log-log plots of $\Delta_{\rm n}$ as a function of 
 $|V-V_{\rm tr}|$, fitted by $|V-V_{\rm tr}|^{\nu}$ with $\nu=5/9$ 
 (tricritical Ising universality class).
 }
 \label{TIM-nu}
\end{figure}

The predictions of perturbed conformal field theory for $\beta$ and
$\nu$ can be checked against numerical computations as follows.
Fixing $U=10.56t$ ($\simeq U_{\rm tr}$), we first give the iDMRG 
results for the CDW order parameter $|\langle m_{\cdw}\rangle|$ as a
function of $V$, cf. Fig.~\ref{TIM-beta}.  Just as in the case of 
the Ising universality class, $|\langle m_{\cdw}\rangle|$ is finite (zero)
for $V>V_{\rm tr}$ ($V<V_{\rm tr}$). The  order parameter $|\langle
m_{\cdw}\rangle|$ now vanishes much more abruptly approaching the
quantum phase transition point from above. Fitting the iDMRG data for
$V>V_{\rm tr}$ to $(V-V_{\rm tr})^{\,\beta}$ with $V_{\rm tr}/t\approx
5.497$  and $\beta=1/24$ works perfectly, see the log-log representation.

In order to verify the field theory prediction for $\nu$ we examine
the $L\to\infty$ extrapolated values of the neutral gap $\Delta_{\rm
  n}$. Increasing $V$($<V_{\rm tr}$) at fixed $U/t=10.56$,
$\Delta_{\rm n}$ is reduced but not linearly as in the Ising case
(cf. Fig.~\ref{Ising-beta}), and closes at $V\approx V_{\rm tr}$
before it becomes finite again for $V>V_{\rm tr}$. 
Again the log-log representation can be used to extract the critical
exponent for $|V-V_{\rm tr}|^\nu$, $\nu=5/9$, for both $V<V_{\rm tr}$
and $V>V_{\rm tr}$, in conformity with the tricritical Ising
universality class. 

\section{Summary}
\label{sec:summary}
To conclude, we have investigated the criticality of the 
1D half-filled extended Hubbard model (EHM) with explicit 
dimerization $\delta$. The BOW-CDW Gaussian transition 
with central charge $c=1$ of the pure EHM gives way to an Ising 
transition with $c=1/2$ at any finite $\delta$.  The Ising transition
line terminates at a tricritical point, which belongs to the
universality class of the tricritical Ising model in two dimensions.
The change of the universality class is verified numerically
by (i)DMRG (see also~\cite{EELF16}). Furthermore, we demonstrate that not only 
the Ising but also the tricritical Ising critical exponents $\beta$ and
$\nu$ can be obtained with high accuracy by simulating the CDW order
parameter and the neutral gap.

\vspace*{1cm}
We thank M. Tsuchiizu for fruitful discussions. The DMRG simulations 
were performed using the ITensor library~\cite{ITensor}. 
This work was supported by Deutsche Forschungsgemeinschaft (Germany), 
SFB 652, project B5 (SE and HF), and by the EPSRC under grant 
EP/N01930X/1 (FHLE).
FL thanks RIKEN for the hospitality sponsored by the IPA program.

%

\end{document}